\documentclass[journal]{IEEEtran}
\usepackage{blindtext}
\usepackage{graphicx}
\usepackage{textcomp}
\usepackage{adjustbox}
\usepackage{xspace}
\makeatletter
\newcommand*{\rom}[1]{\expandafter\@slowromancap\romannumeral #1@}
\newcommand{\degree}{\textdegree\xspace}
\makeatother
\usepackage{atbegshi}
\AtBeginDocument{\AtBeginShipoutNext{\AtBeginShipoutDiscard}}

\usepackage{cite}

   \graphicspath{{./figs/png/}}
	\DeclareGraphicsExtensions{.pdf,.jpeg,.png}

\usepackage[cmex10]{amsmath}

\usepackage[caption=false, font=normalsize, labelfont=sf, textfont=sf]{subfig}

\usepackage{stfloats}

\begin{document}

\title{Modeling and Thermal Metrology of Thermally Isolated MEMS Electrothermal Actuators for Strain Engineering of 2D Materials}

\author{Mounika~Vutukuru,~\IEEEmembership{Member,~IEEE,}
        Jason~W.~Christopher,~\IEEEmembership{Member,~IEEE,}
        Corey~Pollock,~\IEEEmembership{Member, ~IEEE,}
        David~J.~Bishop,~\IEEEmembership{Member,~IEEE,}
        and~Anna~K.~Swan,~\IEEEmembership{Senior Member,~IEEE}}%
        
\thanks{Manuscript received Mon. Day, Year. This work was supported by the National Science Foundation Division of Materials Research under grant number 1411008. DJB is supported by the Engineering Research Centers Program of the National Science Foundation under NSF Cooperative Agreement No. EEC-1647837.}%
\thanks{M. Vutukuru is with the Department of Electrical and Computer Engineering, Boston University, Boston, MA, 02215 USA e-mail: mounikav@bu.edu}%
\thanks{J.W. Christopher is with the Department of Physics, Boston University, Boston, MA, 02215 USA}%
\thanks{C. Pollock is with the Division of Materials Science and Engineering, Boston University, Boston, MA, 02215 USA}%
\thanks{D. J. Bishop is with the Department of Electrical and Computer Engineering and Department of Physics, Boston University, Boston, MA, 02215 USA}%
\thanks{A.K. Swan is with the Department of Electrical and Computer Engineering and Department of Physics, Boston University, Boston, MA, 02215 USA}%

\markboth{Journal of Microelectromechanical Systems,~Vol.~X, No.~X, MONTH~YEAR}%
{Shell \MakeLowercase{\textit{et al.}}: Bare Demo of IEEEtran.cls for Journals}

\maketitle

\begin{abstract}

We present electrothermal microelectromechanical (MEMS) actuators as a practical platform for straining 2D materials. The advantages of the electrothermal actuator is its high output force and displacement for low input voltage, but its drawback is that it is actuated by generating high amounts of heat. It is crucial to mitigate the high temperatures generated during actuation for reliable 2D material strain device implementation. Here, we implement a chevron actuator design that incorporates a thermal isolation stage in order to avoid heating the 2D material from the high temperatures generated during the actuation. By comparing experiment and simulation, we ensure our design does not compromise output force and displacement, while keeping the 2D material strain device stage cool. We also provide a simple analytical model useful for quickly evaluating different thermal isolation stage designs . 

\end{abstract}


\section{Introduction}

2D materials such as graphene and molybdenum disulphide demonstrate tunable material properties under mechanical deformation such as tensile strain, showing great potential for utilization in electronics \cite{Guinea2012}. While their bulk material counterparts, such as bulk silicon, break at nominal strains of less than 2\% \cite{Roldan2015}, atomically thin crystals are capable of withstanding reversible strains up to 20\% without degradation in material quality \cite{Lee2008,Guo2018}. Existing methodologies for strain engineering 2D materials typically include manipulation of the underlying substrate such as bending \cite{Conley2013} or elongating \cite{He2013} flexible substrates, piezoelectric stretching of substrates by applying a gate voltage \cite{Hui2013}, or using differential pressures on membranes over wells \cite{Kitt2013,Lloyd2016}. These techniques pose difficulties in implementation and in integration into commercial electronics. Microelectromechanical systems (MEMS), on the other hand, possess all the functionalities of existing techniques for employing strain with the added advantages of being finely controllable, easily integrated into electronic devices due to their established commercial manufacturing process. MEMS actuators that operate by electrocapacitive or electrothermal action provide displacements in-plane suitable for straining 2D materials. Electrocapacitive “comb-drive” actuators consist of a series of interdigitated fingers acting as microcapacitors, and they are popularly utilized for their zero DC power usage and hence low operating temperatures. However, the actuation voltages run high for electrocapacitive actuators (\textgreater 20V) and the resulting force is only on the order of $ \mu N$, all while taking up significant chip area \cite{Ye1998}. Electrothermal actuators, such as chevron actuators, are a strain platform with high output force (on the order of $ mN $) and displacement with low operating voltages and device footprint \cite{Sinclair2000}. The viability of this platform has already been investigated for strain engineering of metal thin films \cite{Saleh2015} and carbon nanotubes \cite{Zhu2006}. A significant drawback to electrothermal actuators is the generation of heat as means to produce displacement, with actuator temperatures that can swing from room temperature to $ \sim $800 \textdegree C \cite{Que1999}. To use chevron actuators for 2D strain engineering, it is crucial to include thermal management of the 2D films as mechanical and structural properties of the materials are altered with temperature. 

\begin{figure}[!h]
	\centering
	\vspace{-15pt}
\subfloat[Standard Device SD]{%
	\includegraphics[width=1.5in]{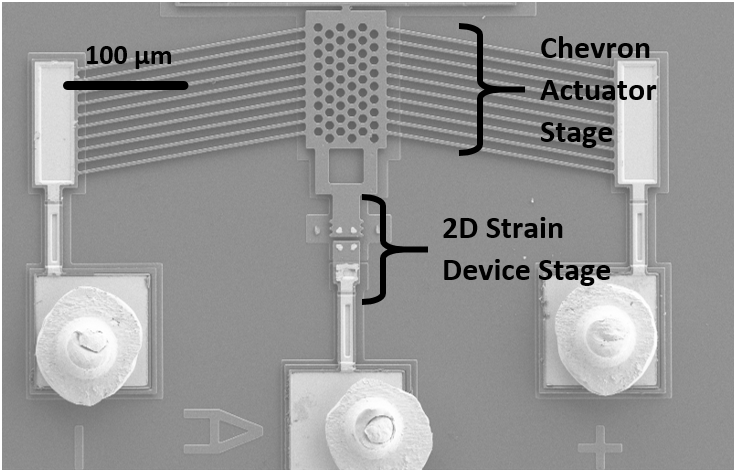}}
\label{1a}\hfill
\subfloat[Thermal Isolation Stage (TIS) Device]{%
	\includegraphics[width=1.8in]{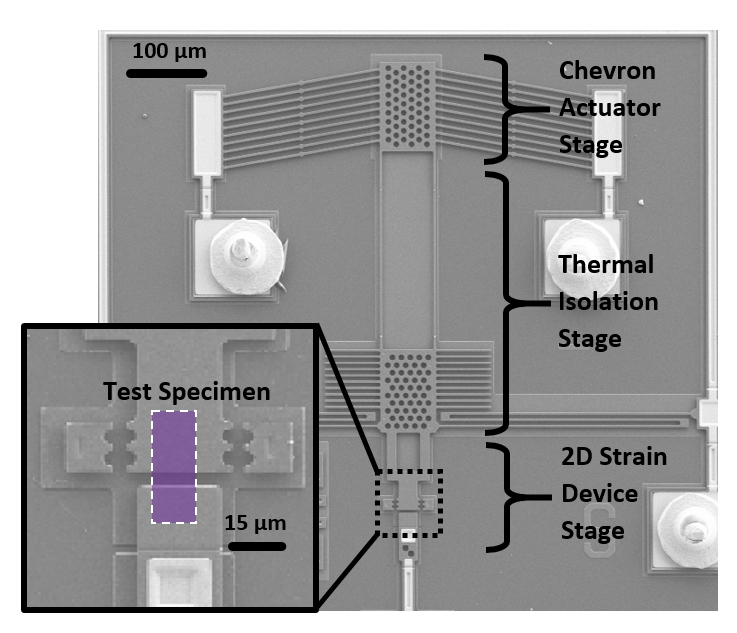}}
\label{1b}\\
\caption{SEM images of Standard Device (SD) and Thermal Isolation Stage (TIS)  Device.  In fig 1 b), the 2D strain device stage is isolated from the heat generated by the chevron actuator by the thermal isolation stage. The inset shows the 2D material Strain Device Stage with a fictitious 2D film in purple spanning the gap between the anchored stage and the moving stage. The scalloped edges provides a visual Vernier.}
\label{SEM}

\end{figure}

 Here we present a chevron actuator device design with a thermal isolation stage (TIS) in order to reduce the temperature at the 2D material strain device stage without sacrificing on output displacement and force. The TIS device is compared with a standard device (SD) without the thermal isolation stage. We have improved the thermal management significantly compared to previous thermal management designs with a less than 10 \textdegree C increase at the 2D strain device stage at the highest actuation voltage while maintaining the high output force ($ \sim $4 $ mN $) and displacement ($ \sim $2.5 $\mu$m) of our standard device. 
 
 Fig \ref{SEM} shows SEM images of chevron type electrothermal microactutors with and without the thermal isolation stage. The SD, in Fig \ref{SEM}a, consists of the chevron actuator stage and 2D strain device stage. Fig \ref{SEM}b shows the TIS device. Our design combines thermal resistor tethers, such as those used by Saleh \textit{et. al.} \cite{Saleh2015}, and heat sink beams \cite{Zhu2006,Qin2013,Zhu2015} to provide thermal management. Previous work using electrothermal actuators for tensile testing of materials deal with the problem of temperature control by thermal loading, which decreases the output displacement/force of the actuator \cite{Qin2013,Abbas2012}. However, our thermal isolation stage successfully prevents the 2D strain device stage from heating without compromising the actuation. 


\section{Device Architecture}
\begin{figure}[!h]
	\centering
	\includegraphics[width=2.5in]{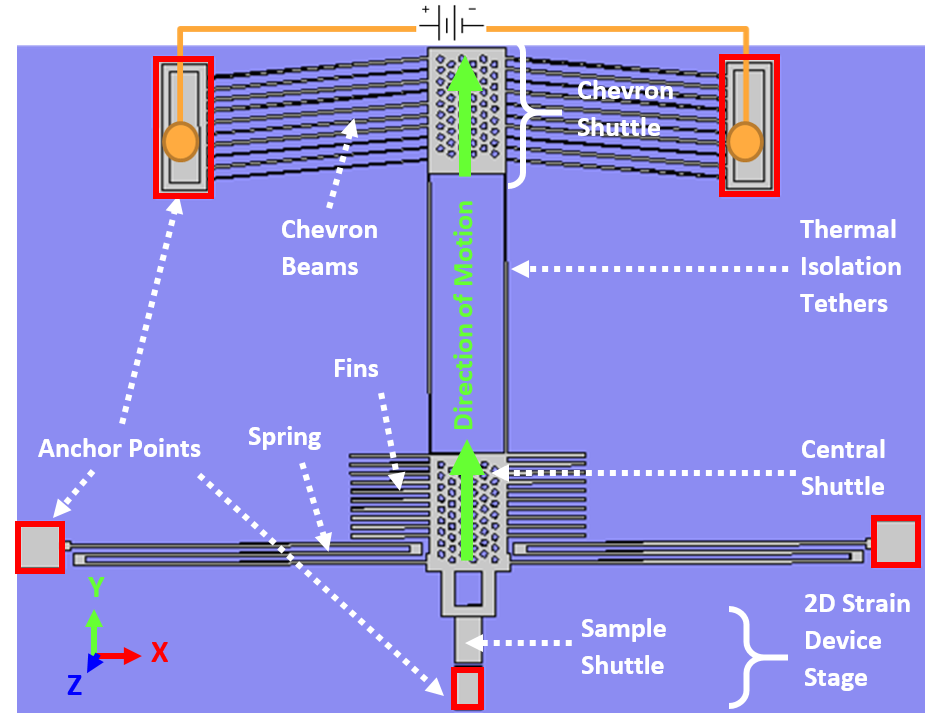}
	\caption{COMSOL Multiphysics model of the TIS device with the different components labeled. All components are free to move except the anchored parts (boxed in red). The joule-heated chevron actuator stage (top) shows the electrical contact points (yellow circles). The green arrows indicates the indicating direction of motion of the actuator. The purple background is the Nitride substrate which is assumed to remain at room temperature under actuation.  }
	\label{COMSOL_TID}
\end{figure}

Our MEMS devices are fabricated by the well-established PolyMUMPS process \cite{Cowen2013}, comprised of three layers of highly degenerately doped, surface micromachined polysilicon on a silicon nitride substrate, interceded with sacrificial oxide layers. On dissolution of the oxide layers in an HF bath, two polysilicon layers are free to move and are raised off the underlying substrate by 2 $\mu m$. The standard chevron actuator design is made of 10 pairs of released polysilicon beams, angled like chevrons, connected to a movable central shuttle. The two sets of chevron beams are anchored at either end to the substrate, where gold contact-pads provide electrical contact. When a voltage is applied across these terminals, the current passing through the beams generates joule heating due to the resistivity of polysilicon. Free to move, these beams undergo thermal expansion due to the ohmic heating, displacing the whole suspended device laterally from its equilibrium position. The angle of the chevron beams, here at 10\degree from the horizontal, directs the majority  of the co-planar motion in the y-direction along the substrate \cite{Que1999}. One end of the central shuttle is attached to the chevron shuttle by two parallel thermal isolation tethers. Finally, the other end of the central shuttle has a 2D strain device stage, where a suspended 2D film can be reliably strained between the moving actuator and another fixed anchor point. Fig \ref{COMSOL_TID} is a CAD model of our TIS device showing the essential features.

\section{Multiphysics Finite Element Modeling and Analysis}

When designing these devices, finite-element modeling is a useful tool to understand the displacement, current, and thermal profile response to a voltage input to the device. Finite-element modeling of both device designs were done to capture the integrated electrical, thermal and structural changes of device under actuation. We use COMSOL Multiphysics to simulate the full geometry of the devices, accessing the Joule heating and thermal expansion modules. Fig. \ref{COMSOL_TID} shows the COMSOL Multiphysics model of the TIS device. 

Structurally, the anchor points shown in the model are the fixed constraints for the devices, while the suspended device is allowed to nonlinearly expand through a temperature-dependent coefficient of thermal expansion. Thermal boundary conditions on the devices allow for thermal conduction through the polysilicon, and via air to the silicon nitride substrate which is constrained to room temperature. Conduction through air and silicon is dominant over convection and radiation at the relevant temperatures \cite{Geisberger2003}. Due to the short air column under the suspended device, conduction through air is very efficient at removing heat at higher temperatures. In contrast, the conduction through polysilicon is less efficient at higher temperatures. Detailed discussion of the heat transfer mechanism through the TIS device is presented in Sec. \rom{7}.

Electrically, an electric potential is applied across the chevron actuator between 0 to 10 V. The electrical conduction through the device is modeled with a temperature-dependent linearized resistivity, using reference resistivities given by PolyMUMPS fabrication processing with each run. The accuracy of our model relies on the temperature-dependent material properties that dictate the electro-thermo-mechanical physics in the device. Furthermore, for polysilicon, these material properties are related to dopant concentration and temperature \cite{McConnell2001,Geisberger2003}, and are nonlinear within the typical actuation range of our devices. Table \rom{1} in the Supplementary Material details the material properties used in the simulation.

The resulting model closely captures the displacement and IV characteristics of the fabricated devices. Our model estimated the maximum displacement for 0.4 W (10 V) of input power for the SD and the TIS device to be $ 2.7 \mu m $ and $ 2.6 \mu m $, respectively. Hence, the improved TIS design is comparable in both displacement and power intake to the original device, from which we can conclude that the thermal isolation stage does not significantly diminish the motion of the shuttle. Complete displacement and IV characteristics for the SD and TIS can be found in Fig. 1 in the Supplementary Material.

\section{Measurements}

To measure the temperature of the microstructures, we used two thermal metrology methods. IR thermometry provides a large-area view of the devices which is useful for distinguishing regions of high temperature in a rapid and qualitative way. However, IR thermometry lacks the temperature range and spatial resolution to resolve smaller areas on the devices, such as chevron beams or thermal isolation tethers. Raman thermometry is valid through the whole temperature range and has high spatial resolution, but assessing temperatures over large areas is time-consuming. The large-area overview and distinct hotspots of the device are easily identifiable with the use of IR while the Raman can be used to determine temperatures of smaller components, such as the chevron actuator beams. Hence, the two methods work in a complementary fashion to fully capture the thermal profile of our devices. 

\subsection{Infrared Thermometry Measurements}

In order to provide a large-area overview of the temperatures in our devices, we actuated the devices using an Optotherm IS640 thermal imaging camera with 5 $\mu m$ resolution, giving a field of view of 3.2 x 2.4 $ mm $ at a working distance o 20 $ mm $. The thermal camera, equipped with an amorphous Si microbolometer detector, operates within the long-wavelength infrared range from 7-14 $\mu m$. Emissivity for the suspended polysilicon microstructures were calibrated by recording the thermal image at different stage temperature using a Peltier thermal stage. The emissivity of the polysilicon for our microstructures was measured to be e = 0.4. This falls in the range of possible spectral emissivities calculated for silicon of 0.4 to 0.71 \cite{Sato1967,Abedrabbo1998,Zhang2000}, and is taken to be constant within the operating temperature range of this device.

\begin{figure}[!h]
	\centering
	\includegraphics[width=3in]{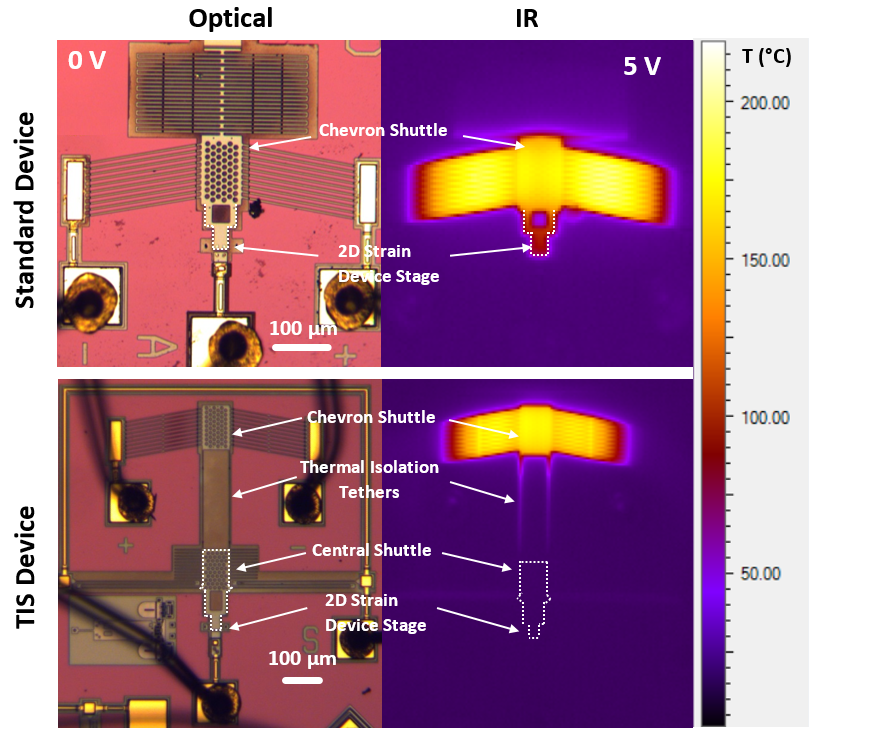}
	\caption{Thermal images of the activated SD and TIS devices.  Left column:    Optical images of un-activated devices (0 V). Right column:   IR thermographs of the activated devices (5V). Temperature is denoted by the color gradient scale from 0 - 230 \textdegree C. The displacement at 5 V is XX um. }
	\label{IR_TID}
\end{figure}

Fig. \ref{IR_TID} shows the devices under the thermal imaging camera when actuated to 0.14 $ W $ (5 V). It is clear that without the thermal isolation stage, the 2D strain device stage is heated to 50\% of the hottest chevron beam temperature, as seen in the SD images. On the other hand, the 2D strain device stage on the TIS device seemingly remains at room temperature for the same actuation conditions. 

\begin{figure}[!h]
	\centering
	\includegraphics[width=3in]{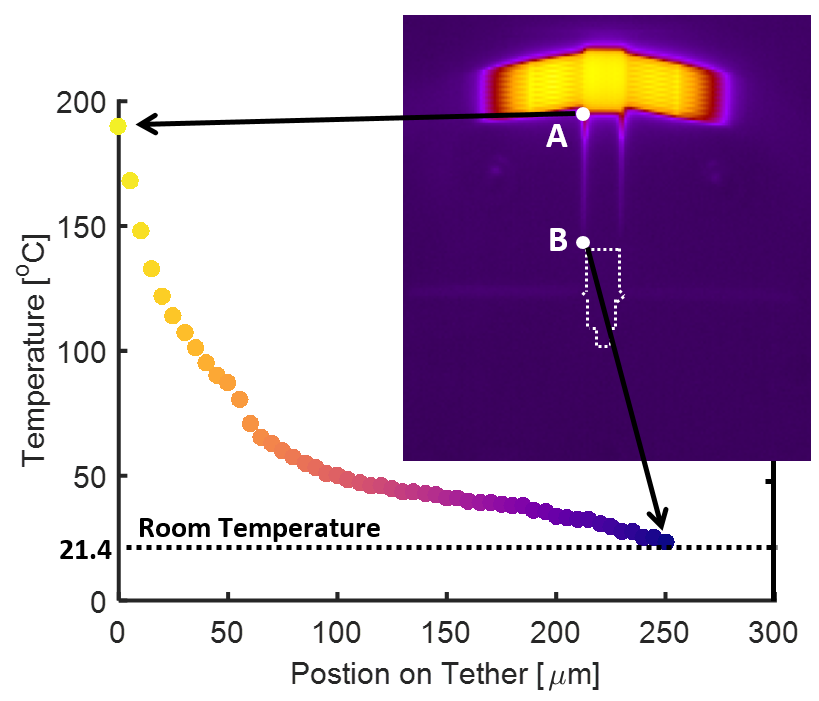}
	\caption{Wide-view IR temperature measurements along a tether. From Point A at the start of the tether near the chevron shuttle to point B near the 2D strain device shuttle, the temperature profile is mapped for the TIS actuated to 0.14 W (5 V). We see that the temperature near the shuttle starts at ~180 \textdegree C  (point A)  and dissipates to room temperature (21.4 \textdegree C) at point B. Color scale  same as in Fig 3.}
	\label{IR_Tether}
\end{figure}

It is clear from the IR thermal profile that the thermal isolation stage is efficient in removing the heat generated by the chevron actuator stage in the TIS device. For 0.14 $ W $ input power, the temperature along the thermal isolation tethers drops from 190 \textdegree C at point A to 21.4 \textdegree C (room temperature) at point B, as shown in Fig. \ref{IR_Tether}. The temperature decrease is nonlinear and drops by $ \sim40\% $ in the first 10th of the thermal isolation tether. This temperature drop is mainly due to conduction through air and not the conduction through the beam, as discussion in section \rom{5}. The thermally isolated 2D strain device stage (outlined with a dashed white line) is indistinguishable from the background in the thermal image indicating it remains at $ \sim $RT.

\subsection{Raman Thermometry Measurements}
Raman spectroscopy is a well-established as a thermal metrology tool for polysilicon microdevices \cite{Serrano2006,Sarua2006,Kearney2006}. It has high spatial resolution, comparable to the minimum device dimensions, as well as high temperature accuracy. In polysilicon, the Raman active mode is $\omega_{0} \sim $520 cm$^{-1}$ which linearly red-shifts with increase in temperature within the typical actuation range of the MEMS actuators. The temperature change $ \Delta T $ has been calibrated to be at a rate of approximately $ C_{Stokes} = $ -0.0232 cm$^{-1}$/ \textdegree C \cite{Abel2007_2} so that $ \Delta T = \Delta\omega/C_{Stokes} $, where $ \Delta\omega $ is the shift of the silicon peak from $ \omega_{0} $.


The Stokes peak position of this Raman mode is also sensitive to stresses in the material. However, free-standing, fully flexible MEMS actuators grown by low-stress LPCVD techniques have been shown to have negligible stress-induced bias on the peak shift \cite{Abel2007}. Effects of internal stresses on the coefficient of Stokes shift are therefore justifiably neglected in calculating temperature. Raman measurements were conducted in-situ on the devices under actuation using a confocal microRaman system (Renishaw Inc). The devices were illuminated using an Argon ion laser, $ \lambda = $514.5 nm excitation, using an Mitutoyo 100x objective (0.58 $\mu m$ beam waist). An irradiation power of 1 $ mW $ was used, which did not measurably optically heat the sample. The scattered signal is detected on a thermoelectrically cooled CCD detector. The spectral pixel resolution is 1 $cm^{-1}$/pixel, while the measurement resolution was two orders of magnitude better \cite{Srikar2003}. 20 spectra were taken for each data point, yielding a consistent photoelectron count of $ 8\times10^{3}$. The standard deviation on each data point was less than 0.01 cm$ ^{-1} $, hence the error bars are not visible in the graph. Fig. 2 in the Supplementary Material demonstrates the effect of temperature on the Stokes peak of the polysilicon as measured from the center of a chevron beam.

\subsection{Temperature Profile Comparison}

\begin{figure}[!h]
	\centering
	\includegraphics[width=3.5in]{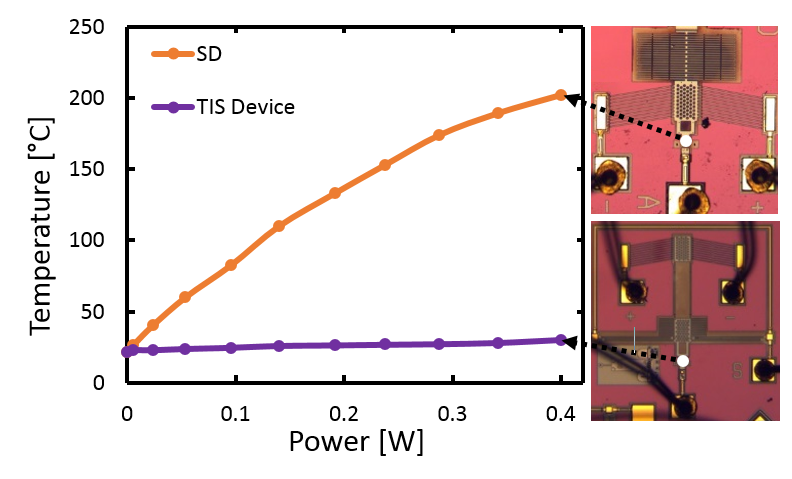}
	\caption{Raman thermometry of the SD and TIS device devices at the center of the 2D strain device stage. SD device: The temperature of the 2D strain device stage increases by 180 C under device actuation.  TIS device: The temperature of the 2D strain device barely increases due to the addition of the thermal isolation stage.}
	\label{TIDvsSD}
\end{figure}

\begin{figure}[!h]
	\centering
	\includegraphics[width=3.5in]{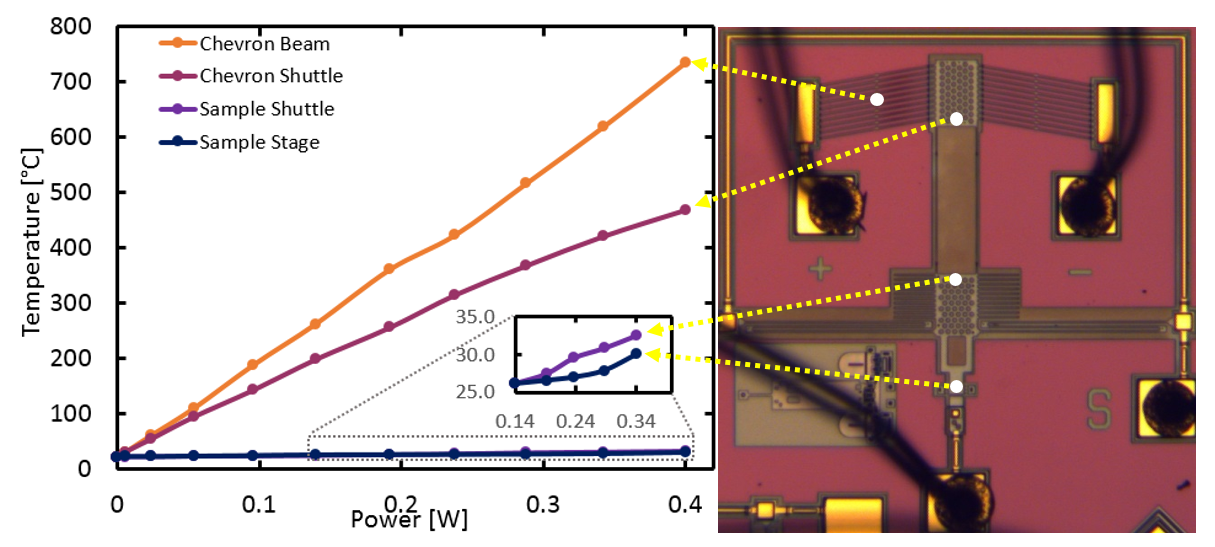}
	\caption{Temperature breakdown for the different stages of the TIS device. At 0.4W (10 V) input power, the highest temperatures are generated at the chevron beams, around 730 \textdegree C. The chevron shuttle at full actuation has a maximum temperature of around 470 \textdegree C. However, temperatures at the sample shuttle and 2D strain device stage rise to a maximum temperature of about 32
	\textdegree C and 30 \textdegree C, respectively. Thus, the thermal isolation stage successfully circumvents the high temperatures generated by the chevron actuator stage during device actuation.}
	\label{TIDStages}
\end{figure}

In order to quantify how well the TIS device works for mitigating large operating temperatures, we take a closer look at the area of interest- the 2D strain device stage of the microactuator. The IR image in Fig. \ref{IR_TID} shows that the 2D strain device stage of the SD experiences high temperatures due to proximity to the actuated chevron beams. Thermal dissipation mechanisms are inefficient over the short range of the device. The 2D strain device stage of the TIS device, on the other hand, is thermally protected from the chevron actuator stage by the thermal isolation stage. We compared the temperature range of the center of the 2D strain device stage through Raman thermometry, Fig. \ref{TIDvsSD}. Over the entire input power range, the 2D strain device stage of the TIS device increases in temperature by less than 10 \textdegree C, reaching a maximum temperature of 30 \textdegree C. By comparison, the 2D strain device stage of the SD increases in temperature by more than 180 \textdegree C, reaching a maximum temperature of 202 \textdegree C. Such temperatures in the region of a 2D film to be tested will affect its mechanical, thermal and/or optical properties. Raman thermometry of multiple points of the TIS device is shown in \ref{TIDStages}. While the 2D strain device stage is thermally isolated, the chevron beams and chevron shuttle reach maximum temperatures of 736 \textdegree C and 469 \textdegree C respectively.

Comparison of Raman and IR thermometries and finite-element simulation of the temperature of the fifth chevron beam show a strong agreement. Details of this comparison can be found in Fig. 3 in the Supplementary Material. Experimental measurements from IR were limited to 0.19 W (6 V) to the device to keep within the calibrated temperature range of the Optotherm IS640 camera. We attribute the slight discrepancy between experimental and simulated data to uncertainty in the temperature-dependent polysilicon material parameters. 

\section{One-dimensional heat transfer and Lumped Thermal Circuit Analysis}

The goal of a thermal isolation stage is to avoid a temperature increase at the 2D strain device stage without thermally loading the chevron shuttle significantly, which would decrease the actuator motion a given power. In this section, we begin with a discussion of the effectiveness of our thermal isolation stage design, and then discuss how the thermal loading on the actuator due to this added stage is negligible. As the core functionality of electrothermal actuators, these devices generate high amounts of heat, particularly at their hottest points on the chevron beams. Our improved design capitalizes on the high output force/displacement of the standard design while simultaneously mitigating high temperature a the sample end. The added thermal isolation stage in our optimized design acts as a voltage divider circuit with shunts to prevent the heat from reaching the 2D strain device stage. We use one-dimensional heat transfer analysis in a simple lumped circuit model. This analysis provides back-of-the-envelope calculations for designing thermal isolation stages under desired constraints, e.g., reducing die space. The thermal circuit model is useful for developing an intuition for the efficiency  and quickly assessing the heat transfer of different designs, since full device simulations using FEA as seen in Sec. \rom{3} are time consuming. 

We begin with considering the mechanisms of one-dimensional heat transfer. Thermal convection and radiation are assessed to be negligible in the temperature range considered, compared to thermal conduction through air and polysilicon \cite{Geisberger2003}. Heat flux $ \overrightarrow{q} $ due to conduction in a material is proportional to the temperature difference $ \Delta T $. The heat transfer along an element in one dimension can then be written as:

\begin{equation}\label{Fourier's Law}
q = -k\frac{A}{\Delta x} \Delta T
\end{equation}

where $ k $ is the material's thermal conductivity, $ A $ is the area normal to heat flow and $ \Delta x $ is length of heat flow. Using the analogue to electrical circuits, we consider the temperature gradient to be analogous with the voltage difference and heat flow to be analogous with electrical current. Thus, we can set up an Ohm's Law relation for a thermal circuit:

\begin{equation}\label{Thermal Ohm's Law}
\Delta T = q\cdot R_{thermal}
\end{equation}

The thermal resistance $ R_{thermal}  $ can therefore be calculated by:

\begin{equation}\label{Thermal Resistance}
R_{thermal} = \frac{\Delta x}{k\cdot A}
\end{equation}

\begin{figure}[!h]
	\centering
	\includegraphics[width=3.5in]{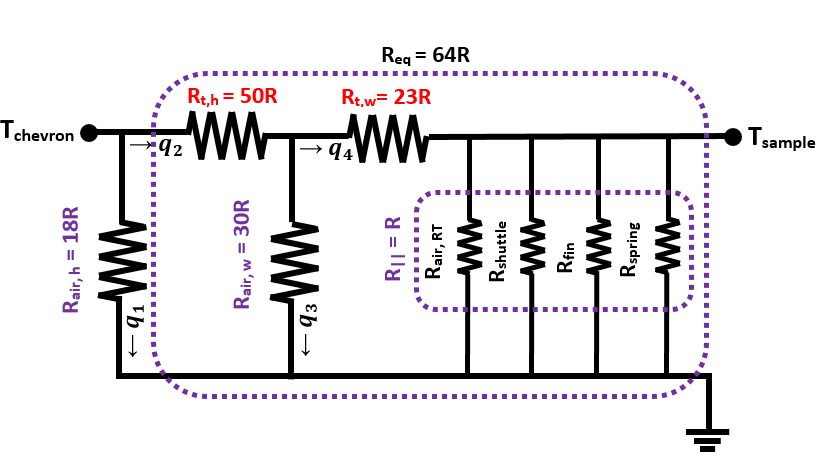}
	\caption{TIS device as a lumped circuit thermal divider. The tether resistors $R_{t,h}$ and $R_{t,w}$ in red conducts heat through the Si tethers. $R_{air,h}$, $R_{air,w}$ and $R_{air,RT}$ in purple conduct heat from the tether through the air-gap  to the substrate (ground), likewise for central shuttle and  resistances ($R_{shuttle}$, $R_{fin}$).  $R_{spring}$ conducts heat through the spring and air. The heat currents are denoted by ${q_{i}}$.
	}
	\label{CircuitAnalysis}
\end{figure}

We simplify our device by modeling components as thermal resistors as seen in Fig. \ref{CircuitAnalysis}. We use Eqn. (\ref{Thermal Resistance}) to calculate resistance values. There are two conduction pathways: through the polysilicon, and through air between the suspended MEMS device and the silicon nitride substrate kept at room temperature. At high temperatures, conduction through air dominates over conduction through polysilicon, while the opposite is true for lower temperatures. To analyze the model, we use only three temperatures which determine the conductivities. The maximum temperature of the chevron shuttle, $ T_{chevron} $, for the first portion of the tether is used to calculate $ R_{t,h} $ and $ R_{air,h} $. At the halfway point of the tether, an intermediate "warm" temperature $ T_{w} $ is used for $ R_{t,w} $ and $ R_{air,w} $. Finally, room temperature at the end of the tether and onwards, $ T_{RT} $ is used for the remaining resistances in the circuit model. We use IR thermometry to guide the temperature choices. We determined that the temperature $ T $ varies as function of position $ x $ along the tether such that $ T \propto x^{-1/2} $ works reasonably well. Hence, we assign the temperature in the middle of the tether as roughly 1/4 of $ T_{chevron} $. For simplicity, the rest of the thermal isolation stage used room temperature values for thermal conductivities. The temperature of interest at the 2D strain device stage is $ T_{sample} $. A summary of our circuit analysis variables and resistances, details of the temperature along the tether and $ k_{si} $ and $ k_{air} $  as a function of input power and temperature can be found in Table \rom{2} and Figs. 4 and 5 in the Supplementary Material, respectively.

As seen Fig. \ref{CircuitAnalysis}, an equivalent resistance $ R_{||} $ can be used to replace the four mutually parallel resistances $ R_{air, RT} $, $ R_{shuttle} $, $ R_{fin} $, $ R_{spring} $. We can also make an equivalent resistance $ R_{eq} $ for our circuit as indicated by the dashed line in purple in Fig. \ref{CircuitAnalysis}. We want the overall resistance of the TIS to be low enough so that the TIS does not significantly thermally load the actuator. In order to shunt the heat through air and prevent it from reaching the 2D strain device stage, we want $ q_{1} > q_{2} $ or $ R_{air} < R_{eq} $. Likewise, we want $ q_3 > q_4 $. Finally, $ R_{||} $ should be small compared to $ R_{t,w} $ to keep $ T_{sample} $ as close room temperature as possible.

Our aim is to find the temperature of the sample $ T_{sample} $ in relation with temperature generated at the chevron shuttle $ T_{chevron} $. Using the previously defined resistances, these temperatures can be related as:

\begin{equation}
\Delta T_{sample} = \frac{R_{||}R_{air,w}}{R_{eq}(R_{t,w}+R_{||}+R_{air,w})} \Delta T_{chevron}
\end{equation}

Details on the derivation can be found in Section \rom{1} in the Supplementary Material. Using the calculated resistance values, we can calculate:

\begin{equation}
\Delta T_{sample} = 0.8\% \Delta T_{chevron}
\end{equation}

According to the simple lumped circuit model, the 2D strain device stage increases in temperature by a mere $ \sim $1\% of the temperature generated at the chevron shuttle due to the effectiveness of our thermal isolation stage. At the maximum input power, 1\% of $ T_{chevron} $ would give a temperature at $ T_{sample} $ = 25 \textdegree C. This gets reasonably close to the measured value of 30 \textdegree C and is useful for a first pass of designing the thermal isolations stage. We followed this analysis for another input power 0.14 W (5 V) of our device, which results in $ T_{sample} = 1.5\% T_{chevron} $. This results in $ T_{sample} $ = 24 \textdegree C, again very close to the measured value of 25.7 \textdegree C, showing that our modeling provides a good rule of thumb for these types thermal isolation stages. Details of the calculations for the second input power as well the general trend of the TIS effectiveness and $ T_{sample} $ as it varies with input power can be found in Table \rom{3} and Fig. 7 in the Supplementary Material.

\section{Conclusion}

The lumped circuit model has clearly been simplified many times over and for more serious design reviews of the TIS several tests with different temperatures is required. For example, for a more accurate model, better choices of temperature can be determined iteratively. However, as we have shown, our model is sufficient for a first pass at analyzing the effectiveness of the thermal isolation stage. 

Recent works on using chevron type actuators for straining materials note that the trade-off for effective thermal isolation of the test specimen is a decrease in output force and displacement \cite{Qin2013}. We have presented a thermally isolated device that does not compromise on the output displacement (\texttildelow 2.5 $ \mu m $) achieved by our standard design. Despite the added thermal isolation stage, our design remains compact, occupying a 800 $ \mu $m x 800 $ \mu $m footprint. As determined by our thermal metrology techniques, we observe a temperature increase of $ \leq $ 10 \textdegree  C from room temperature on our 2D strain device stage at actuation temperatures of 735 \textdegree C. Finally, we have developed a simple lumped thermal circuit model for analyzing the thermal isolation stage. This model can be useful for future works on optimizing the TIS device, using back-of-the-envelop calculations instead of time-consuming simulations. For example, to achieve even better thermal isolation of the 2D strain device stage, the tether resistance could be increased by making the tether length longer.


%


\section*{Acknowledgment}
M. Vutukuru would like to thank the BUnano Fellowship for their support.

\bibliographystyle{ieeetran}
\bibliography{mybibfile}

\vfill

\begin{IEEEbiography}{Mounika Vutukuru} recieved the B.S. degree in electrical engineering with a concentration in nanotechnology and the B.A. degree in physics from Boston University in 2015. She is currently working toward the Ph.D. degree in electrical engineering at Boston University. Her research interests include strain engineering of 2D materials using microelectromechanical systems for the purposes of probing unique physics and prototyping novel electronic devices. 
\end{IEEEbiography}	

\begin{IEEEbiography}{Jason W. Christopher} received the B.S. degree in Physics and Electrical Engineering and Computer Science from Massachusetts Institute of Technology in 2005, and spent several years working as an electrical engineer in Silicon Valley, most notably as a technical lead for the Trackpad Team at Apple, Inc. He then received the M.S. and Ph.D. degrees in Physics from Boston University in 2017 and 2018, respectively, where he studied mechanisms for controlling strain in 2D materials and measured the effect of strain on material properties using Raman and Photoluminescence spectroscopies. Currently, he is applying his expertise in electronics, measurement, and statistical inference to develop unobtrusive methods for measuring patient vital signs and creating models to predict patient wellness at Myia Labs, Inc.
\end{IEEEbiography}

\begin{IEEEbiography}{Corey Pollock} recieved the B.S. degree in mechanical engineering from the University of Washington in 2010 and the M.S. degree in mechanical engineering from Boston University in 2017, where he is currently pursuing the Ph.D. degree in mechanical engineering. He is currently a member of the Dr. D. Bishop Group, Boston University. His research involves microelectromechanical systems technology, specifically studying micromirrors and smart lighting applications. 
\end{IEEEbiography}
	
\begin{IEEEbiography}{David J. Bishop} (M'11) received the B.S. degree in physics from Syracuse University in 1973, and the M.S. and Ph.D. degrees in physics from Cornell University in 1977 and 1978, respectively. He joined the AT\&T-Bell Laboratories Bell Labs in 1978 as a Post-Doctoral Member of Staff and became a Member of the Technical Staff in 1979. In 1988, he was made a Distinguished Member of the Technical Staff and later he was promoted as the Department Head, Bell Laboratories. He was the President of Government Research and Security Solutions with Bell Labs, Lucent Technologies. He was the Chief Technology Officer and Chief Operating Officer with LGS, the wholly-owned subsidiary of Alcatel-Lucent dedicated to serving the U.S. federal government market with advanced research and development solutions. He is currently the Head of the Division of Materials Science and Engineering, Boston University, a Professor of Physics, and also a Professor of Electrical Engineering. He is a Bell Labs Fellow and in his previous positions with Lucent served as a Nanotechnology Research VP for Bell Labs, Lucent Technologies, and the President of the New Jersey Nanotechnology Consortium and the Physical Sciences Research VP. He is a member and a fellow of the American Physical Society, and a member of the MRS. He was a recipient of the APS Pake Prize.
\end{IEEEbiography}

\begin{IEEEbiography}{Anna K. Swan}  received a BSc and MS degree in Physics Engineering from Chalmers University, Gothenburg, Sweden, and a Ph.D. degree in Physics at Boston University, Boston, MA, in 1994. Her dissertation topic was the spin-ordering on NiO(100) surfaces using metastable He  scattering, for which she received two student awards, the Nottingham prize and the Morton M. Traum Award. She joined the Solid State Division at Oak Ridge National Laboratory as a Wigner Fellow. and later as a staff member. In 2005, she joined the Electrical and Computer Engineering Department at Boston University where she is currently an Associate Professor and Associate chair of graduate studies. Her research interests  clustered around high-spatial resolution spectroscopy. Currently she is working on 2D materials and their responses to strain and charge using photoluminescence  and  micro-Raman spectroscopy.
\end{IEEEbiography}




%

\end{document}